\documentclass[aps,prd]{revtex4}

\usepackage{amsmath}
\usepackage{amsfonts}
\usepackage{amssymb}
\usepackage{graphicx}

\begin{document}
%-----------------------------------------------
\title{Closed timelike curves and causality violation}
%-----------------------------------------------

\author{Francisco S. N. Lobo}
\email{flobo@cii.fc.ul.pt}\affiliation{Centro de Astronomia
e Astrof\'{\i}sica da Universidade de Lisboa, Campo Grande, Ed. C8
1749-016 Lisboa, Portugal}

\date{\today}

%-----------------------------------------------
\begin{abstract}
%-----------------------------------------------

The conceptual definition and understanding of time, both quantitatively and qualitatively is of the utmost difficulty and importance. As time is incorporated into the proper structure of the fabric of spacetime, it is interesting to note that General Relativity is contaminated with non-trivial geometries which generate {\it closed timelike curves}. A closed timelike curve (CTC) allows time travel, in the sense that an observer that travels on a trajectory in spacetime along this curve, may return to an event before his departure. This fact apparently violates causality, therefore time travel and it's associated paradoxes have to be treated with great caution. The paradoxes fall into two broad groups, namely the {\it consistency paradoxes} and the {\it causal loops}. A great variety of solutions to the Einstein field equations containing CTCs exist and it seems that two particularly notorious features stand out. Solutions with a tipping over of the light cones due to a rotation about a cylindrically symmetric axis and solutions that violate the energy conditions. All these aspects are analyzed in this review paper.

%-----------------------------------------------
\end{abstract}
%-----------------------------------------------

%\pacs{04.50.-h, 04.50.Kd, 04.20.Jb}

%-----------------------------------------------
\maketitle
%-----------------------------------------------

%\newpage

\tableofcontents

\section{Introduction}

Providing an explicit definition of time is an extremely difficult endeavor, although it does seem to be intimately related to change, an idea reflected in Aristotle's famous metaphor: {\it Time is the moving image of Eternity} \cite{Bert_Lobo}. In fact, one may encounter many reflections and philosophical considerations on time over the ages, culminating in Newton's notion of absolute time. Newton stated that time flowed at the same rate for all observers in the Universe. But, in 1905, Einstein changed altogether our notion of time. Time flowed at different rates for different observers, and Minkowski, three years later, formally united the parameters of time and space, giving rise to the notion of a four-dimensional entity, spacetime. Adopting a pragmatic point of view, this assumption seems reasonable, as to measure time a changing configuration of matter is needed, i.e., a swinging pendulum, etc. Change seems to be imperative to have an emergent notion of time.
Therefore, time is empirically related to change. But change can
be considered as a variation or sequence of occurrences \cite{Kluwer}. Thus,
intuitively, a sequence of successive occurrences provides us
with a notion of something that flows, i.e., it provides us with
the notion of {\it time}. Time flows and everything relentlessly
moves along this stream. In Relativity, we can substitute the
above empirical notion of a sequence of occurrences by a sequence
of {\it events}. We idealize the concept of an event to become a
point in space and an instant in time. Following this reasoning of
thought, a sequence of events has a determined {\it temporal
order}. We experimentally verify that specific events occur before
others and not vice-versa. Certain events (effects) are triggered
off by others (causes), providing us with the notion of {\it
causality}.

Thus, the conceptual definition and understanding of time, both
quantitatively and qualitatively is of the utmost difficulty and
importance. Special Relativity provides us with important
quantitative elucidations of the fundamental processes related to
time dilation effects. The General Theory of Relativity (GTR)
provides a deep analysis to effects of time flow in the presence
of strong and weak gravitational fields \cite{Simon}. As time is incorporated into the proper structure of the fabric of spacetime, it is interesting to note that GTR is contaminated with non-trivial geometries which generate {\it closed timelike curves} \cite{Visser,Springer,Kluwer,LLQ-PRD,Tipler-CTCs}. A closed timelike curve (CTC) allows time travel, in the sense that an observer which travels on a trajectory in spacetime along this curve, returns to an event which coincides with the departure. The arrow of time leads forward, as measured locally by the observer, but globally he/she may return to an event in the past. This fact apparently violates causality, opening Pandora's
box and producing time travel paradoxes \cite{Nahin}, throwing a
veil over our understanding of the fundamental nature of Time. The
notion of causality is fundamental in the construction of physical
theories, therefore time travel and it's associated paradoxes have
to be treated with great caution. The paradoxes fall into two
broad groups, namely the {\it consistency paradoxes} and the {\it
causal loops}.

The consistency paradoxes include the classical grandfather
paradox. Imagine traveling into the past and meeting one's
grandfather. Nurturing homicidal tendencies, the time traveler
murders his grandfather, impeding the birth of his father,
therefore making his own birth impossible. In fact, there are many
versions of the grandfather paradox, limited only by one's
imagination. The consistency paradoxes occur whenever
possibilities of changing events in the past arise.
The paradoxes associated to causal loops are related to
self-existing information or objects, trapped in spacetime.
Imagine a time traveler going back to his past, handing his
younger self a manual for the construction of a time machine. The
younger version then constructs the time machine over the years,
and eventually goes back to the past to give the manual to his
younger self. The time machine exists in the future because it was
constructed in the past by the younger version of the time
traveler. The construction of the time machine was possible
because the manual was received from the future. Both parts
considered by themselves are consistent, and the paradox appears
when considered as a whole. One is liable to ask, what is the
origin of the manual, for it apparently surges out of nowhere.
There is a manual never created, nevertheless existing in
spacetime, although there are no causality violations. An
interesting variety of these causal loops was explored by Gott and
Li~\cite{Gott-Li}, where they analyzed the idea of whether there
is anything in the laws of physics that would prevent the Universe
from creating itself. Thus, tracing backwards in time through the
original inflationary state a region of CTCs may be encountered,
giving {\it no} first-cause.

A great variety of solutions to the Einstein Field Equations
(EFEs) containing CTCs exist, but, two particularly notorious
features seem to stand out. Solutions with a tipping over of the
light cones due to a rotation about a cylindrically symmetric
axis; and solutions that violate the Energy Conditions of GTR,
which are fundamental in the singularity theorems and theorems of
classical black hole thermodynamics \cite{hawkingellis}. A great
deal of attention has also been paid to the quantum aspects of
closed timelike curves \cite{quantum1,quantum2,quantum3}.

Throughout this paper, we use the notation $G=c=1$.

\section{Stationary and axisymmetric solutions generating CTCs}

It is interesting to note that the tipping over of light cones seems to be a generic feature of some solutions with a rotating cylindrical symmetry.
The general metric for a stationary, axisymmetric solution with rotation is
given by \cite{Visser,Wald}
\begin{equation}
ds^2=-F(r)\,dt^2+H(r)\,dr^2+L(r)\,d\phi^2+2\,M(r)\,d\phi \,dt+H
(r)\,dz^2  \,,   \label{stationarymetric}
\end{equation}
where $z$ is the distance along the axis of rotation; $\phi$ is the angular coordinate; $r$ is the radial coordinate; and $t$ is the temporal coordinate. The metric components are only functions of the radial coordinate $r$. Note that the determinant, $g={\rm det}(g_{\mu\nu})=-(FL+M^2)H^2$, is Lorentzian provided that $(FL+M^2)>0$.

Due to the periodic nature of the angular coordinate, $\phi$, an
azimuthal curve with $\gamma =\{t={\rm const},r={\rm const},z={\rm
const}\}$ is a closed curve of invariant length $s_{\gamma}^2
\equiv L(r)(2\pi)^2$. If $L(r)$ is negative then the integral
curve with $(t,r,z)$ fixed is a CTC. If $L(r)=0$, then the
azimuthal curve is a closed null curve. Now, consider a null azimuthal curve, not necessarily a geodesic nor closed, in the $(\phi,t)$ plane with $(r,z)$ fixed. The null condition, $ds^2=0$, implies
\begin{equation}
0=-F+2M\dot{\phi}+L\dot{\phi}^2  \,,
\end{equation}
with $\dot{\phi}=d\phi/dt$. Solving the quadratic, we have
\begin{equation}
\frac{d\phi}{dt}=\dot{\phi}=\frac{-M\pm \sqrt{M^2+FL}}{L}  \,.
\end{equation}

\begin{figure}[h]
\centering
\includegraphics[width=2.75in]{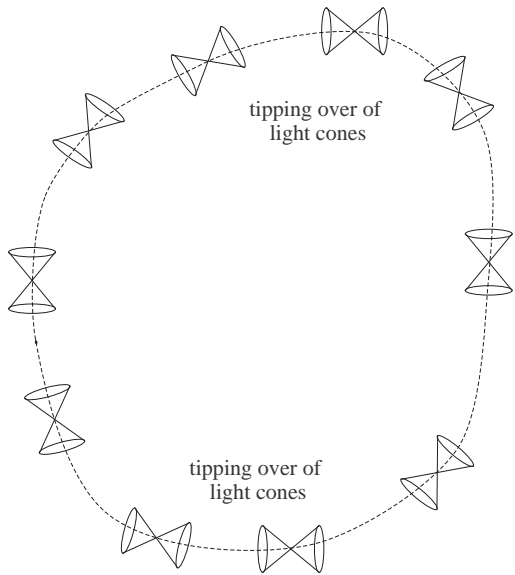}
\caption{The tipping over of light cones, depicted in the figure
is a generic feature of some solutions with a rotating cylindrical
symmetry. The dashed curve represents a closed timelike curve.}
\label{tipcones}
\end{figure}

Due to the Lorentzian signature constraint, $FL+M^2>0$, the
roots are real. If $L(r)<0$ then the light cones are tipped over
sufficiently far to permit a trip to the past. By going once
around the azimuthal direction, the total backward time-jump for a
null curve is
\begin{equation}
\Delta T=\frac{2\pi |L|}{-M+ \sqrt{M^2-F|L|}}   \,.
\end{equation}
If $L(r)<0$ for even a single value of $r$, the chronology-violation region covers the entire spacetime \cite{Visser}. Thus, the tilting of light cones are generic features of spacetimes which contain CTCs, as depicted in Fig. \ref{tipcones}.

The present section is far from making an exhaustive search of all
the EFE solutions generating CTCs with these features, but the
best known spacetimes will be briefly analyzed, namely, the van
Stockum spacetime, the G\"{o}del universe, the spinning cosmic
strings and the Gott two-string time machine, which is a variation
on the theme of the spinning cosmic string.

\subsection{Van Stockum spacetime}

The earliest solution to the EFEs containing CTCs, is probably
that of the van Stockum spacetime, which describes a stationary,
cylindrically symmetric solution of a rapidly rotating
infinite cylinder of dust, surrounded by vacuum. The centrifugal
forces of the dust are balanced by the gravitational attraction.
The metric, assuming the respective symmetries, takes the form of
Eq. (\ref{stationarymetric}), and $t$ is required to be
timelike at $r=0$. The coordinates $(t,r,\phi,z)$ have the
following domain
\begin{equation}
-\infty <t<+\infty, \qquad 0<r<\infty, \qquad 0\leq\phi\leq2\pi,
\qquad  -\infty <z<+\infty   \,.
\end{equation}

\subsubsection{The Interior solution}

The metric for the interior solution $r<R$, where $R$ is the surface of the
cylinder, is given by
\begin{equation}
ds^2=-dt^2+2\omega r^2 d\phi dt+r^2(1-\omega^2
r^2)d\phi^2+\exp(-\omega^2 r^2)(dr^2+dz^2)
\end{equation}
where $\omega$ is the angular velocity of the cylinder. It is
immediate to verify that CTCs arise if $\omega r>1$, i.e., for
$r>1/\omega$ the azimuthal curves with $(t,r,z)$ fixed are CTCs.
The condition $M^2+FL=\omega^2 r^4+r^2(1-\omega^2 r^2)=r^2>0$ is
imposed.

The causality violation region could be eliminated by requiring
that boundary of the cylinder to be at $r=R<1/a$. The interior
solution would then be joined to an exterior solution, which would
be causally well-behaved. The resulting upper bound to the
``velocity'' $\omega R$ would be $1$, although the orbits of the
particles creating the field are timelike for all $r$.

Applying the EFE, the energy density and $4$-velocity of the dust
are given by
\begin{eqnarray}
8\pi \rho=4\omega^2 \exp(\omega^2 r^2) \qquad {\rm and} \qquad
U^\mu=(1,0,0,0) \,,
\end{eqnarray}
respectively. The coordinate system co-rotates with the dust. The
source is simply positive density dust, implying that all of the
energy condition are satisfied.

\subsubsection{The Exterior solution}

Van Stockum developed a procedure which generates an exterior
solution for all $\omega R>0$~\cite{Tipler}. Consider the
following range:

%\medskip

\begin{enumerate}

\item[] $(i)$ $0<\omega R<1/2$.

The exterior solution is given by the following functions
\begin{eqnarray*}
&&H(r)=\exp(-\omega^2 r^2) \left ( r/R \right )^{-2\omega^2 r^2},
\qquad
L(r)=\frac{Rr \sinh(3\varepsilon +\theta)}{2\sinh(2\varepsilon)
\cosh(\varepsilon)}\,,   \\
&&M(r)=\frac{r \sinh(\varepsilon +\theta)}{\sinh(2\varepsilon)},
\qquad
F(r)=\frac{r \sinh(\varepsilon -\theta)}{R\sinh(\varepsilon)}  \,,
\end{eqnarray*}
with
\begin{eqnarray*}
\theta=\theta(r)=(1-4\omega ^2 R^2)^{1/2} \ln \left(r/R \right )
\qquad  {\rm and}  \qquad
\varepsilon =\varepsilon (r)={\rm arctanh} (1-4\omega ^2
R^2)^{1/2}   \,.
\end{eqnarray*}

\item[] $(ii)$ $\omega R = 1/2$.
\begin{eqnarray*}
&&H(r)=\exp^{-1/4} \left ( r/R \right )^{-1/2},
\qquad
L(r)=(Rr/4) \left [3+\ln \left(r/R \right ) \right ],\\
&&M(r)=(r/2) \left [1+\ln \left(r/R \right ) \right ],
\qquad
F(r)=(r/R) \left [1-\ln \left(r/R \right ) \right ] \,.
\end{eqnarray*}

\item[] $(iii)$ $\omega R>1/2$.
\begin{eqnarray*}
&&H(r)=\exp(-\omega^2 r^2) \left (r/R \right )^{-2\omega^2 r^2},
\qquad
L(r)=\frac{Rr \sin(3\beta +\gamma)}{2\sin(2\beta)
\cos(\beta)},\\
&&M(r)=\frac{r \sin(\beta +\gamma)}{\sin(2\beta)},
\qquad
F(r)=\frac{r \sin(\beta -\gamma)}{R\sin(\beta)}  \,,
\end{eqnarray*}
with
\begin{eqnarray*}
\gamma=\gamma(r)=(4\omega^2 R^2-1)^{1/2} \ln \left(r/R \right )
\qquad {\rm and} \qquad
\beta=\beta(r)=\arctan(4\omega^2 R^2-1)^{1/2} \,.
\end{eqnarray*}
\end{enumerate}

As in the interior solution, $FL+M^2=r^2$, so that the metric
signature is Lorentzian for $R\leq r<\infty $.

\subsubsection{Chronology violation region}

\begin{figure}[h]
  \centering
  \includegraphics[width=4.0in]{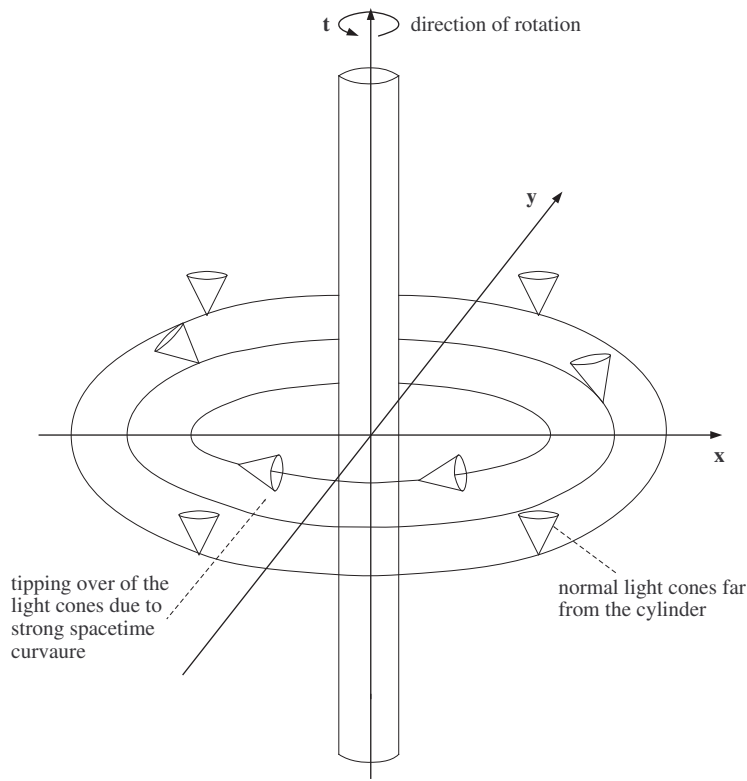}
  \caption[Tipping over of light cones in the van Stockum spacetime]
  {Van Stockum spacetime showing the tipping over of
  light cones close to the cylinder, due to the strong curvature of
  spacetime, which induce closed timelike curves.}\label{fig:stockum}
\end{figure}

One may show that the causality violation is avoided for $\omega R\leq 1/2$, but in the region $\omega R> 1/2$, CTCs appear. The causality violations arise from the sinusoidal factors of the metric components. The first zero of $L(r)$ occurs at
\begin{equation}
r_0=R \left [\frac{\pi-3\arctan(4\omega^2
R^2-1)^{\frac{1}{2}}}{(4\omega^2 R^2-1)^{\frac{1}{2}}} \right ]
\,.
\end{equation}
Thus, causality violation occur in the matter-free space surrounding a
rapidly rotating infinite cylinder, as shown in Figure
\ref{fig:stockum}.
The van Stockum spacetime is not asymptotically flat. But, the
gravitational potential of the cylinder's Newtonian analog also
diverges at radial infinity. Shrinking the cylinder down to a
``ring'' singularity, one ends up with the Kerr solution, which
also has CTCs (The causal structure of the Kerr spacetime has been
extensively analyzed by de Felice and
collaborators~\cite{Felice1,Felice2,Felice3,Felice4,Felice5}).

In summary, the van Stockum solution contains CTC provided $\omega
R>1/2$. The chronology-violating region covers the entire
spacetime. Reactions to the van Stockum solution is that it is
unphysical, as it applies to an infinitely long cylinder and it is
not asymptotically flat.

\subsection{The G\"{o}del Universe}

Kurt G\"{o}del in $1949$ discovered an exact solution to the EFEs
of a uniformly rotating universe containing dust and a nonzero
cosmological constant~\cite{Godel}. The total energy-momentum is
given by
\begin{equation}
T^{\mu\nu}_{\rm
total}=\rho\,U^\mu\,U^\nu-\frac{\Lambda}{8\pi}\,g^{\mu\nu} \,.
\end{equation}
However, the latter may be expressed in terms of a perfect fluid, with rotation, energy density
$\bar{\rho}$ and pressure $\bar{p}$, in a universe with a zero
cosmological constant, i.e.,
\begin{equation}
T^{\mu\nu}_{\rm
total}=(\bar{\rho}+\bar{p})\,U^\mu\,U^\nu+\bar{p}\;g^{\mu\nu} \,,
\end{equation}
with the following definitions
\begin{equation}
\bar{\rho}= \rho+\frac{\Lambda}{8\pi} \qquad {\rm and}\qquad
\bar{p}=-\frac{\Lambda}{8\pi} \,.
\end{equation}

The manifold is $R^4$ and the metric of the G\"{o}del solution is
provided by
\begin{equation}
ds^2=-dt^2-2e^{\sqrt{2}\,\omega
x}\,dtdy-\frac{1}{2}\,e^{2\sqrt{2}\,\omega x}\,dy^2+dx^2+dz^2  \,.
   \label{Godelmetric}
\end{equation}
The four-velocity and the vorticity of the fluid are,
$U^\mu=\delta^{\mu}{}_{0}=(1,0,0,0)$ and
$\omega^\mu=(0,0,0,\omega)$, respectively. The Einstein field
equations provide the following stress-energy scenario:
\begin{equation}
4\pi \rho= \omega^2=-\Lambda \qquad {\rm or}\qquad
\bar{p}=\bar{\rho}=\frac{\omega^2}{8\pi}>0 \,.
\end{equation}
Thus, the null, weak and dominant energy conditions are satisfied, while the dominant energy condition is in the imminence of being violated.

Note that the metric (\ref{Godelmetric}) is the direct sum of the
metrics ${\bf g_1}$ and ${\bf g_2}$. The metric ${\bf g_1}$ is given by
\begin{equation}
ds_1^2=-dt^2-2e^{\sqrt{2}\,\omega
x}\,dtdy-\frac{1}{2}\,e^{2\sqrt{2}\,\omega x}\,dy^2+dx^2  \,,
   \label{Godelmetric1}
\end{equation}
with the manifold ${\cal M}_1=R^3$ defined by the coordinates
$(t,x,y)$. The metric ${\bf g_2}$ is given by $ds^2_2=dz^2$, with the
manifold ${\cal M}_2=R$, defined by the coordinate $z$.

To analyze the causal properties of the solution, it is sufficient
to consider $({\cal M}_1,{\bf g_1})$. Consider a set of
alternative coordinates $(t',r,\phi)$ in $({\cal M}_1,{\bf g_1})$,
in which the rotational symmetry of the solution, around the axis
$r=0$, is manifest and suppressing the irrelevant $z$ coordinate,
defined by~\cite{hawkingellis,Godel}
\begin{eqnarray}
\omega\,y\, e^{\sqrt{2}\,\omega\,x}&=&\sin\phi\;\sinh(2r) \,,
            \nonumber \\
e^{\sqrt{2}\,\omega\,x}&=&\cosh(2r)+\cos\phi\;\sinh(2r)  \,,
            \nonumber   \\
\tan\left[\left(\phi+\omega\,t-\sqrt{2}\,t'\right)/2\right]&=&
e^{-2r}\,\tan(\phi/2)
            \nonumber     \,,
\end{eqnarray}
so that the metric (\ref{Godelmetric1}) takes the form
\begin{equation}
ds^2=2w^{-2} \left[-dt'^2+dr^2-(\sinh ^4r-\sinh ^2r)\,d\phi
^2+2(\sqrt{2})\sinh ^2r \,d\phi \,dt \right]  \,.
\end{equation}
Moving away from the axis, the light cones open out and tilt in
the $\phi$-direction. The azimuthal curves with $\gamma =\{t={\rm
const},r={\rm const},z={\rm const}\}$ are CTCs if the condition
$r>\ln (1+\sqrt{2})$ is satisfied.

It is interesting to note that in the G\"{o}del spacetime, closed
timelike curves are not geodesics. However, Novello and
Rebou\c{c}as \cite{Nov-Reb} discovered a new generalized solution
of the G\"{o}del metric, of a shear-free nonexpanding rotating
fluid, in which successive concentric causal and noncausal regions
exist, with closed timelike curves which are geodesics. A complete
study of geodesic motion in G\"{o}del's universe, using the method
of the effective potential was further explored by Novello {\it et
al}~\cite{Nov-Soar}. Much interest has been aroused in time travel
in the G\"{o}del spacetime, from which we may mention the analysis
of the geodesical and non-geodesical motions considered by
Pfarr~\cite{Pfarr} and Malament~\cite{Malament1,Malament2}.

\subsection{Gott Cosmic String time machine}

\subsubsection{Gravitational field of a Cosmic String}

The string spacetime is assumed to be static and cylindrically
symmetric, with the string lying along the axis of symmetry. The
most general static, cylindrically symmetric metric has the form
% line element
\begin{equation}
ds^2=-e^{2\nu(\rho)}\,dt^2+e^{2\,\lambda(\rho)}\,(d\rho^2+dz^2)
+e^{2\Phi(\rho)}\,d\phi^2
\,,
\end{equation}
where $\nu$, $\Phi$ and $\lambda$ are functions of $\rho$. $\phi
=0$ and $\phi =2\pi$ are identified.

Suppose that the string has a uniform density $\epsilon >0$, out
to some cylindrical radius $\rho_0$. The end results will prove to
be independent of $\rho_0$, so that the string's transverse
dimensions may be reduced to zero, yielding an unambiguous exact
exterior metric for the string.

The stress-energy tensor of the string, in an orthonormal frame,
is given by
\begin{equation}
T_{\hat{t}\hat{t}} =-T_{\hat{z}\hat{z}} =\epsilon
\end{equation}
and all the other components are equal to zero, for $\rho< \rho_0$~\cite{VilenkinPRD,Linet,Hiscock-string}. The resulting
EFEs are given by:
%    G(up,dn)   [1, 1]
\begin{eqnarray}
&&-e^{-2\lambda}\,\left[\Phi''+(\Phi')^2+\lambda''
\right] =8\pi \,\epsilon  \,,  \label{stringEFEtt}  \\
&&e^{-2\lambda}\,\left(\lambda'\,\Phi'+\nu'\,\lambda'+\lambda'\,\Phi'
\right) =0  \,,  \label{stringEFErr}  \\
&&e^{-2\lambda}\,\left[\lambda''+\nu''+(\nu')^2
\right] =0  \,,  \label{stringEFEthetatheta}  \\
&&e^{-2\lambda}\,\left[-\lambda'\,\Phi'-\nu'\,\lambda'+\Phi''
+(\Phi')^2+\nu''+(\nu')^2+\nu'\,\Phi' \right] =-8\pi\,\epsilon \,,
    \label{stringEFEzz}
\end{eqnarray}
where the prime denotes a derivative with respect to $\rho$. These
are non-linear equations for the metric functions, and are easily
solved in the case of the uniform density string. Conservation of
the stress-energy, $T^{\beta}{}_{\alpha ;\beta}=0$, yields
\begin{equation}
(\nu'+\lambda') \epsilon =0  \,.
\end{equation}
This implies that through Eq. (\ref{stringEFEthetatheta}), $\nu$
and $\lambda$ are constant, and may be set to zero by an
appropriate rescaling of the coordinates $t$, $\rho$ and $z$.
Equation (\ref{stringEFErr}) is then satisfied automatically and
eqs. (\ref{stringEFEtt}) and (\ref{stringEFEzz}) become identical,
i.e.,
\begin{equation}
\Phi''+(\Phi')^2 =-8 \pi \epsilon  \,.
\end{equation}

Substituting $R=e^{\Phi}$, i.e., $g_{\phi\phi}=R^2$, yields
\begin{equation}
R=A \cos(\rho/\bar{\rho})+B \sin(\rho/\bar{\rho}) \,,
\end{equation}
where $\bar{\rho}=(8\pi\epsilon)^{-1/2}$. The metric on the axis
will be flat, i.e., no cone singularity, if $A=0$ and $B=
\bar{\rho}$. Thus, the interior metric of a uniform-density string
is then given by
\begin{equation}
ds^2=-dt^2+d\rho^2+\frac {1}{8\pi\epsilon}\,\sin^2\left(\sqrt
{8\pi \epsilon}\;\rho\right)\,d\phi^2+dz^2   \,.
\end{equation}

The exterior metric for the string spacetime must be a static,
cylindrically symmetric, vacuum solution of the EFEs. The most
general solution, discovered by Levi-Civita~\cite{Levi-Civita} is
given
\begin{equation}
ds^2=-r^{2m}dT^2+r^{-2m}\left
[r^{2m^2}(dr^2+dZ^2)+a^2r^2d\phi^{2}\right ] \label{extstring}\,,
\end{equation}
where $m$ and $a$ are freely chosen constants. The string is Lorentz
invariant in the $z$-direction. Requiring that the metric
(\ref{extstring}) be Lorentz invariant in the $z$-direction
restricts the values of $m$, namely, $m=0$ and
$m=2$~\cite{Kramer-exact}.

One may now join the interior and exterior metrics together along
the surface of the string at $\rho= \rho_0$ and $r= r_0$. The
Darmois-Israel junction conditions require that the intrinsic
metrics induced on the junction surface by the interior and
exterior metrics be identical, and that the discontinuity in the
extrinsic curvature of the surface be related to the surface
stress-energy. Consider the $m=0$ flat exterior case.

The intrinsic metric can then be matched by requiring $t=T$, $z=Z$
and $g_{\phi\phi}^{+}=g_{\phi\phi}^{-}$. The latter condition
provides
\begin{equation}
ar_0=\bar{\rho}\;\sin(\rho_0/ \bar{\rho})  \,.
\label{stringintrinsic}
\end{equation}
Calculating the extrinsic curvature tensors and equating them to each other, so as to have no surface stress-energy present, one obtains the following
relation
\begin{equation}
a^2=\frac{\bar{\rho}^2}{\bar{\rho}^2+r_{0}^2}  \,.
\end{equation}
Combining this with the intrinsic metric constraint, Eq.
(\ref{stringintrinsic}), to eliminate $r_0$, yields
\begin{equation}
a=\cos(\rho_0/ \bar{\rho}) \,.  \label{def:a}
\end{equation}
The exterior metric of the string is then given by Eq.
(\ref{extstring}) with $m=0$, and $a$ given by Eq. (\ref{def:a}).

The concept of a mass per unit length for a cylindrically
symmetric source in general relativity is not unambiguously
defined, unlike the case of spherical symmetry. For a static,
cylindrically symmetric spacetime, a useful simple definition is
to integrate the energy-density, $\epsilon$ over the proper volume
of the source, i.e., the string.

The mass per unit length, or linear energy-density, is given by
\begin{equation}
\mu=\int_0^{\rho_0} \int_0^{2\pi}\epsilon \bar{\rho}\,\sin(\rho/
\bar{\rho})\;d\phi \,d\rho=2\pi \,\epsilon \,\bar{\rho}\,^2 \left
[1-\cos(\rho/ \bar{\rho})\right ]  \,,
\end{equation}
which, taking into account $\bar{\rho}=(8\pi\epsilon)^{-1/2}$,
reduces to
\begin{equation}
4\mu=1-\cos(\rho_0/ \bar{\rho})  \,.
\end{equation}
Thus, the exact exterior metric is given by
\begin{equation}
ds^2=-dt^2+dr^2+\left (1-4\,\mu\right )^2\,r^2\,d\theta^2+dz^2 \,,
\label{exactextstring}
\end{equation}
which will be used below in the Gott cosmic string spacetime.

\subsubsection{Gott Cosmic String spacetime}

An extremely elegant model of a time-machine was constructed by Gott \cite{GottCTC}. The Gott time-machine is an exact solution of the EFE for the general case of two moving straight cosmic strings that do not intersect \cite{GottCTC}. This solution produces CTCs even though they do not violate the WEC, have no singularities and event horizons, and are not topologically multiply-connected as the wormhole solution (see below). The appearance of CTCs relies solely on the gravitational lens effect and the relativity of simultaneity. We follow the analysis of Ref. \cite{GottCTC} closely throughout this section.

The exterior metric of a straight cosmic string is given by Eq.
(\ref{exactextstring}). The geometry of a $t={\rm const}$, $z={\rm
const}$ section of this solution is that of a cone with an angle
deficit $D=8\pi \mu$ in the exterior (vacuum) region. Applying a
new coordinate $\phi'=(1-4\mu)\phi$, the exterior metric becomes
\begin{equation}
ds^2=-dt^2+dr^2+r^2\,d\phi'\,^2+dz^2  \,,
\end{equation}
where $0\leq \phi'<(1-4\mu)\,2\pi$. The above metric is the
metric for Minkowski space in cylindrical coordinates where a
wedge of angle deficit $D=8\pi \mu$ is missing, and points with
coordinates $(r, \phi'=0,z,t)$ and $(r, \phi'=2\pi-8\pi \mu,z,t)$
are identified.

Now, the static solution for two parallel cosmic strings separated by a
distance $2d$ is constructed in the following manner. Consider the metric
(\ref{exactextstring}), by replacing the angular and radial coordinates, $\phi'$ and $r$, respectively by the Cartesian coordinates, $x=r \sin(\phi'+4\pi \mu)$ and $y=r \cos(\phi'+4\pi \mu)+d$. This reduces the metric to $ds^2=-dt^2+dx^2+dy^2+dz^2$, with the following restrictions:
\begin{equation}
x^2+(y-d)^2\geq r_b^2  \,,  \qquad   |x|\geq (y-d) \tan(4\pi \mu)
\,,
\end{equation}
and the points with $x=\pm (y-d) \tan(4\pi \mu)$ are identified.

\begin{figure}[h]
  \centering
  \includegraphics[width=3.4in]{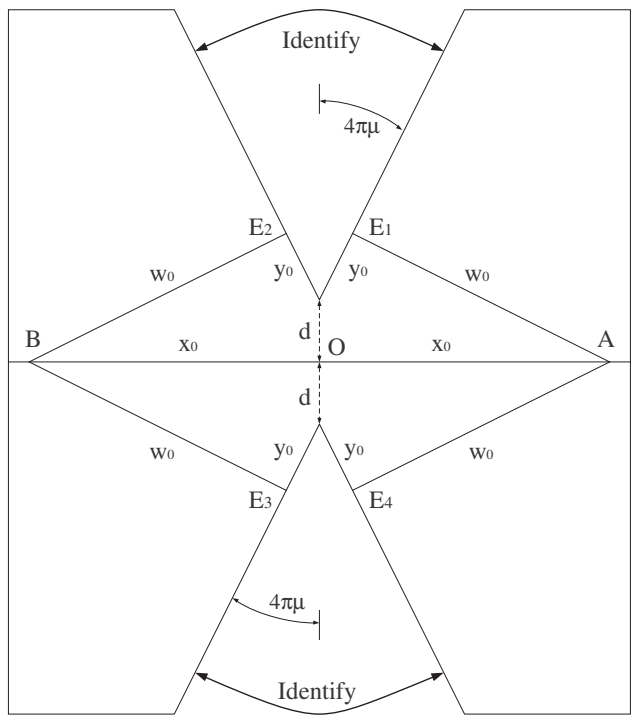}
  \caption[Gott two-string spacetime in the $(x,y)$ plane]
  {Two-parallel string static spacetime on the $(x,y)$ plane}
  \label{GottTM}
\end{figure}

The $3$-surface $y=0$, has the metric $ds^2=-dt^2+dx^2+dz^2$ with
zero intrinsic and extrinsic curvature, as it is part of a
$(3+1)$-dimensional Minkowski spacetime. It is thus possible to
produce a mirror-image copy of the region $y\geq 0$, including the
interior solution, by joining it along the three-surface $y=0$.
This second solution lies in the $y\leq 0$ region. The two copies
obey all the matching conditions along the surface $y=0$, because
the latter is a $(2+1)$-dimensional Minkowskian spacetime with
zero intrinsic and extrinsic curvature. See Figure \ref{GottTM}
for details.

Consider now two observers $A$ and $B$ at rest with respect
to the cosmic strings with world lines given by
\begin{equation}
x_A ^{\mu}=(t_A,x_A,y_A,z_A)=(t,x_0,0,0) \qquad {\rm and} \qquad
x_B ^{\mu}=(t_B,x_B,y_B,z_B)=(t,-x_0,0,0)  \,,
\end{equation}
respectively. It is possible to prove that observer $B$ sees three images of the observer $A$ \cite{Gott-lensing}. The central image is from a
geodesic passing through the origin $0$, the two outer images,
which are displaced from the central image by an angle $\Delta
\theta=4\pi \mu$ on each side, represent geodesics that pass
through events $E_1-E_2$ and $E_3-E_4$. Note that $E_1$ and $E_2$
are identified, as are $E_3$ and $E_4$.

Considering the following trigonometric relationship
\begin{equation}
w^2=(x_0-y_0 \sin(4\pi \mu))^2+(d+y_0 \cos (4\pi \mu))^2  \,,
\end{equation}
it is simple to verify that the value of $y_0$ to minimize $w_0$
is $y_0=x_0 \sin(4\pi \mu)-d \cos (4\pi \mu)$. Thus, we have
$w_0<x_0$ if $d<y_0$, and the light beam going through $0$ with a
gravitational lensing time delay between the two images of $\Delta
t=2(x_0-w_0)$. Note that if a light beam traversing through $E_1-E_2$ can
beat a light beam traveling through $0$, then so can  a spaceship
traveling at a high enough velocity, $\beta_r <1$, relative to
the string. The spaceship connects two events in the $y=0$
$(2+1)$-dimensional Minkowski spacetime with a spacelike
separation.

Let the spaceship begin at $A$ and end at $B$, given by the
following events
\begin{equation}
E_i=(-\beta_r^{-1} w_0,x_0,0,0) \qquad {\rm and} \qquad
E_f=(\beta_r^{-1} w_0,-x_0,0,0)   \,,
\end{equation}
respectively. The time for the spaceship to traverse from $E_i$,
through $E_1-E_2$ to $E_f$ is $t=2\beta_r w_0$. The separation of
$E_i$ and $-E_f$ is spacelike providing that $x_0^2-\beta_r^{-2}
w_0^{2}>0$, which can always be verified for high enough
$\beta_r<1$, for $w_0<x_0$.

The following step is to give the $y\geq 0$ solution a boost with
velocity $\beta_s$ in the $+x$-direction via a Lorentz
transformation such that $E_i$ and $E_f$ become simultaneous in
the laboratory frame. The velocity for the simultaneity to occur
is $\beta_s=w_0 \beta_r^{-1} x_0^{-1}$.
Analogously, we give the $y\leq0$ solution a boost with velocity
$\beta_s$ in the $-x$-direction. The two solutions $y\geq0$ and
$y\leq0$ may still be matched together because the Lorentz
transformations do not alter the fact that the boundary surface
$t=0$ in each solution is still a $(2+1)$-dimensional Minkowskian
spacetime with zero intrinsic and extrinsic curvature.

The spaceship goes from $E_i$ through $E_1-E_2$ and arrives at
$E_f$, which is simultaneous in the laboratory frame. By symmetry,
the spaceship travels in the opposite direction past the
oppositely moving string through $E_3-E_4$ and arrives back at
event $E_i$, which is also simultaneous with $E_f$ in the
laboratory frame. The spaceship has completed a CTC, as it
encircles the two parallel cosmic strings as they pass each other
in a sense opposite to that of the strings' relative motion. In principle, it is also possible to find a reference frame in which the spaceship
arrives at $E_i$ before it's departure.

The events in the laboratory frame have the following coordinates:
$E_{i,L}=(0,\gamma_s^{-1} x_0,0,0)$ and $E_{f,L}=(0,-\gamma_s^{-1}
x_0,0,0)$ with $\gamma_s^{2}=\frac{x_0 ^2}{x_0 ^2-\beta_r^{-2}
w_0^{2}}$ and since $\beta_r<1$, we have
\begin{equation}
\gamma_s^{2}> \frac{x_0 ^2}{x_0 ^2-w_0^{2}}=\frac{y_0 ^2}{x_0
^2-d^{2}}  \,,
\end{equation}
or
\begin{equation}
\gamma_s^{2}> \frac{(\sin(4\pi\mu))^{-2}}{1-\frac{2d}{x_0
\tan(4\pi\mu)}-\frac{d^2}{x_0^2}}  \,.
\end{equation}
Considering the following approximations, $x_0 \gg d$, we have
\begin{equation}
\gamma_s>(\sin(4\pi\mu))^{-1}   \,,
\end{equation}
or simply
\begin{equation}
\beta_s>\cos(4\pi\mu)  \,.
\end{equation}
For $\mu=10^{-6}$ expected for grand unified cosmic strings, we
have $\gamma_s>8\times 10^4$ in order to produce CTCs.

In the laboratory frame it is clear how the CTC is created. The
$E_1-E_2$ and $E_3-E_4$ identifications allow the particle to
effectively travel backward in time twice in the laboratory frame.
The identifications of $E_1-E_2$ and $E_3-E_4$ is equivalent to
having a complete Minkowski spacetime  without the missing wedges
where instantaneous, tachyonic, travel in the string rest frames
between $E_1$ and $E_2$, $E_3$ and $E_4$, is possible.

It is also interesting to verify whether the CTCs in the Gott
solution appear at some particular moment, i.e., when the strings
approach each other's neighborhood, or if they already pre-exist,
i.e., they intersect any spacelike hypersurface. These questions
are particularly important in view of Hawking's Chronology
Protection Conjecture \cite{hawking}. This conjecture states that
the laws of physics prevent the creation of CTCs. If correct, then
the solutions of the EFE which admit CTCs are either unrealistic
or are solutions in which the CTCs are pre-existing, so that the
time -machine is not created by dynamical processes. Amos Ori
proved that in Gott's spacetime, CTCs intersect every $t={\rm
const}$ hypersurface \cite{Ori:CTCstring}, so that it is not a
counter-example to the Chronology Protection Conjecture.

The global structure of the Gott spacetime was further explored by
Cutler~\cite{Cutler-CTC}, and it was shown that the closed
timelike curves are confined to a certain region of the spacetime,
and that the spacetime contains complete spacelike and achronal
hypersurfaces from which the causality violating regions evolve.
Grant also examined the global structure of the two-string
spacetime and found that away from the strings, the space is
identical to a generalized Misner space~\cite{Grant-CTC}. The
vacuum expectation value of the energy-momentum tensor for a
conformally coupled scalar field was then calculated on the
respective generalized Misner space, which was found to diverge
weakly on the chronology horizon, but diverge strongly on the
polarized hypersurfaces. Thus, the back reaction due to the
divergent behavior around the polarized hypersurfaces are
expected to radically alter the structure of spacetime, before
quantum gravitational effects become important, suggesting that
Hawking's chronology protection conjecture holds for spaces with a
noncompactly generated chronology horizon. Soon after,
Laurence~\cite{Laurence-CTC} showed that the region containing
CTCs in Gott's two-string spacetime is identical to the regions of
the generalized Misner space found by Grant, and constructed a
family of isometries between both Gott's and Grant's regions. This
result was used to argue that the slowly diverging vacuum
polarization at the chronology horizon of the Grant space carries
over without change to the Gott space. Furthermore, it was shown
that the Gott time machine is unphysical in nature, for such an
acausal behavior cannot be realized by physical and timelike
sources \cite{Deser,Deser2,Deser3,Farhi,Farhi2}.

\subsection{Spinning Cosmic String}

Consider an infinitely long straight string that lies and spins
around the $z$-axis. The symmetries are analogous to the van
Stockum spacetime, but the asymptotic behavior is different
\cite{Visser,Jensen}. We restrict the analysis to an infinitely
long straight string, with a delta-function source confined to the
$z$-axis. It is characterized by a mass per unit length, $\mu$; a
tension, $\tau$, and an angular momentum per unit length, $J$. For
cosmic strings, the mass per unit length is equal to the tension,
$\mu=\tau$.

In cylindrical coordinates the metric takes the following form
\begin{equation}
ds^2=-\left [d(t+4J\varphi) \right
]^2+dr^2+(1-4\mu)^2\,r^2\;d\varphi^2+dz^2 \,,
\end{equation}
with the following coordinate range
\begin{equation}
-\infty <t<+\infty,  \qquad   0<r<\infty,  \qquad   0\leq\varphi
\leq 2\pi,  \qquad  -\infty <z<+\infty   \,.
\end{equation}

Adopting a new set of coordinates
\begin{equation}
\overline{t}=t+4J\varphi \\
\overline{\varphi}=(1-4J)\varphi  \,,
\end{equation}
the metric may be rewritten as
\begin{equation}
ds^2=-d\overline{t}^2+dr^2+r^2d\overline{\varphi}^2+dz^2  \,,
\end{equation}
with a new coordinate range
\begin{equation}
-\infty <t<+\infty,  \qquad  0<r<\infty,  \qquad  0\leq
\overline{\varphi} \leq (1-4\mu)2\pi,  \qquad   -\infty
<z<+\infty \,,
\end{equation}
subject to the following identifications
\begin{equation}
(\overline{t},r,\overline{\varphi},z)\equiv \left[\overline{t}+8\pi
J,r,\overline{\varphi}+2\pi(1-4\mu ),z\right]  \,.
\end{equation}

Outside the core $r=0$, the metric is locally flat, i.e., the
Riemann tensor is zero. The geometry is that of flat Minkowski
spacetime subject to a somewhat peculiar set of identifications.
On traveling once around the the string, one sees that the
spatial slices are "missing" a wedge of angle $8\pi \mu$, which
defines the deficit angle $\Delta \theta=8\pi \mu$.
On traveling once around the string, one undergoes a backward
time-jump of
\begin{equation}
\Delta \overline{t}=8\pi J   \,.
\end{equation}

Consider an azimuthal curve, i.e., an integral curve of $\varphi$.
Closed timelike curves appear whenever
\begin{equation}
r<\frac{4J}{1-4\mu}  \,.
\end{equation}
These CTCs can be deformed to cover the entire spacetime,
consequently, the chronology-violating region covers the entire
manifold.

\section{Solutions violating the Energy Conditions}

The traditional manner of solving the EFEs, $G_{\mu \nu}=8\pi
T_{\mu \nu}$, consists in considering a plausible stress-energy
tensor, $T_{\mu \nu}$, and finding the geometrical structure,
$G_{\mu\nu}$. But one can run the EFE in the reverse direction by
imposing an exotic metric $g_{\mu\nu}$, and eventually finding the
matter source for the respective geometry. In this fashion,
solutions violating the energy conditions have been obtained.
Adopting the reverse philosophy, solutions such as traversable
wormholes, the warp drive, the Krasnikov tube and the Ori-Soen
spacetime have been obtained. These solutions violate the energy
conditions and with simple manipulations generate CTCs.

\subsection{Conversion of traversable wormholes into time machines}

Much interest has been aroused in traversable wormholes since the
classical article by Morris and Thorne \cite{Morris}. A wormhole
is a hypothetical tunnel which connects different regions in
spacetime. These solutions are multiply-connected and probably
involve a topology change, which by itself is a problematic issue.

Consider the following spherically symmetric and static wormhole
solution
\begin{equation}
ds^2=-e ^{2\Phi(r)} \,dt^2+\frac{dr^2}{1- b(r)/r}+r^2 \,(d\theta
^2+\sin ^2{\theta} \, d\phi ^2) \,, \label{metricwormhole}
\end{equation}
where $\Phi(r)$ and $b(r)$ are arbitrary functions of the radial
coordinate $r$. $\Phi(r)$ is denoted the redshift function, for it
is related to the gravitational redshift, and $b(r)$ is denoted
the shape function, because as can be shown by embedding diagrams,
it determines the shape of the wormhole \cite{Morris}. The
coordinate $r$ is non-monotonic in that it decreases from
$+\infty$ to a minimum value $r_0$, representing the location of
the throat of the wormhole, where $b(r_0)=r_0$, and then it
increases from $r_0$ to $+\infty$.

A fundamental property of a wormhole is that a flaring out
condition of the throat, given by $(b-b^{\prime}r)/b^{2}>0$, is
imposed \cite{Morris}, and at the throat
$b(r_{0})=r=r_{0}$, the condition $b^{\prime}(r_{0})<1$ is imposed
to have wormhole solutions. It is precisely these restrictions
that impose the NEC violation in classical general relativity.
Another condition that needs to be satisfied is $1-b(r)/r>0$. For
the wormhole to be traversable, one must demand that there are no
horizons present, which are identified as the surfaces with
$e^{2\Phi}\rightarrow0$, so that $\Phi(r)$ must be finite
everywhere.

Several candidates have been proposed in the literature, amongst which we refer to solutions in higher dimensions, for instance in Einstein-Gauss-Bonnet theory \cite{EGB1,EGB2}, wormholes on the brane \cite{braneWH1}; solutions in Brans-Dicke theory \cite{Brans-Dicke}; wormholes constructed in $f(R)$ gravity \cite{Lobo:2009ip}; wormhole solutions in semi-classical gravity (see Ref. \cite{Garattini:2007ff} and references therein); exact wormhole solutions using a more systematic geometric approach were found \cite{Boehmer:2007rm}; wormhole solutions and thin shells \cite{thinshells}; geometries supported by equations of state responsible for the cosmic acceleration \cite{phantomWH}; spherical wormholes were also formulated as an initial value problem with the throat serving as an initial value surface \cite{MontelongoGarcia:2009zz}; solutions in conformal Weyl gravity were found \cite{Weylgrav}, and thin accretion disk observational signatures were also explored \cite{Harko:2008vy}, etc (see Refs. \cite{Lemos:2003jb,Lobo:2007zb} for more details and \cite{Lobo:2007zb} for a recent review).

One of the most fascinating aspects of wormholes is their apparent
ease in generating CTCs \cite{mty,wh-timemachine}. There are several ways to
generate a time machine using multiple wormholes \cite{Visser},
but a manipulation of a single wormhole seems to be the simplest
way \cite{mty,visser-babyuniv}. The basic idea is to create a time
shift between both mouths. This is done invoking the time dilation
effects in special relativity or in general relativity, i.e., one
may consider the analogue of the twin paradox, in which the mouths
are moving one with respect to the other, or simply the case in
which one of the mouths is placed in a strong gravitational field.

\begin{figure}[h]
\centering
  \includegraphics[width=2.6in]{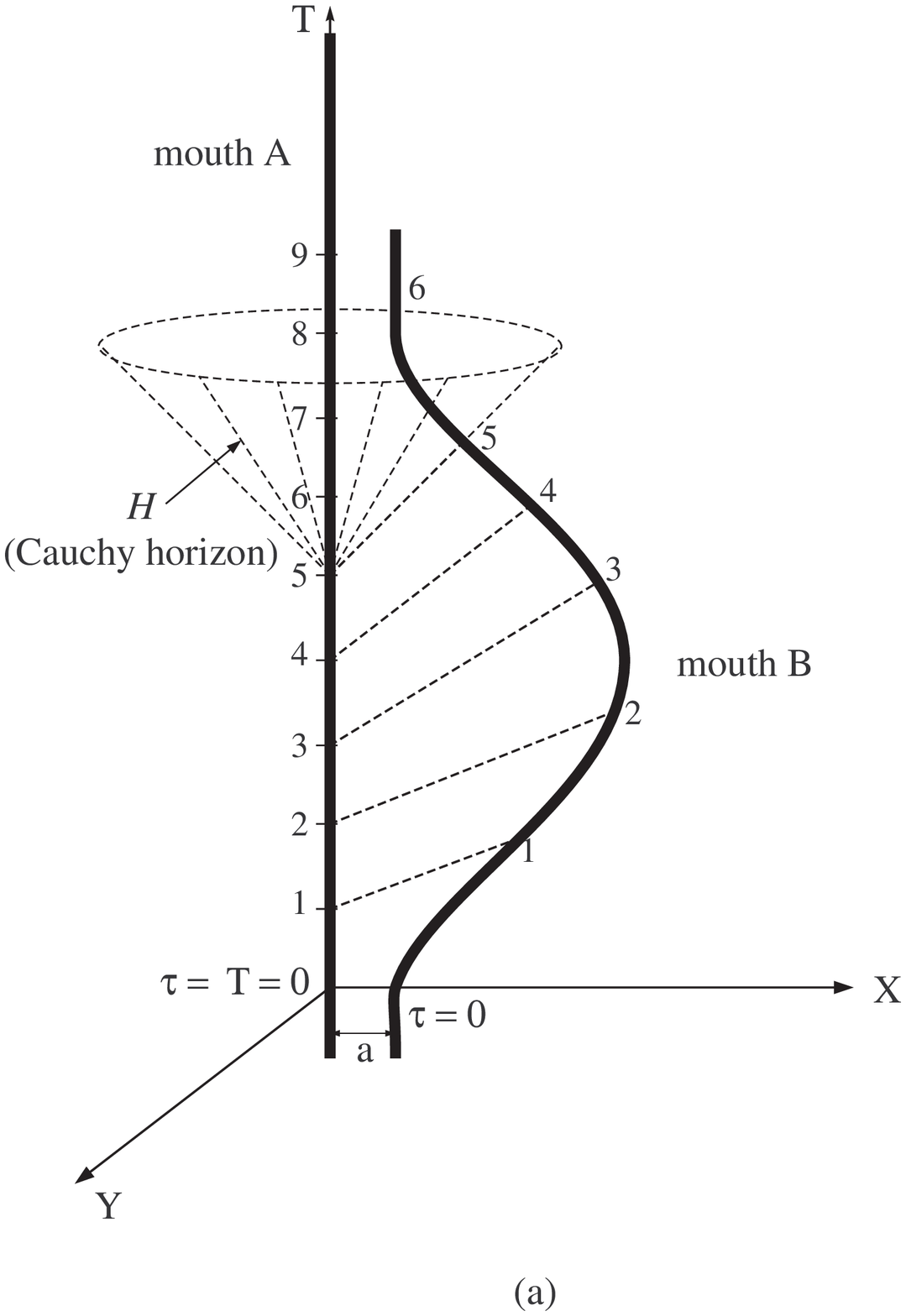}
  \hspace{0.4in}
  \includegraphics[width=2.4in]{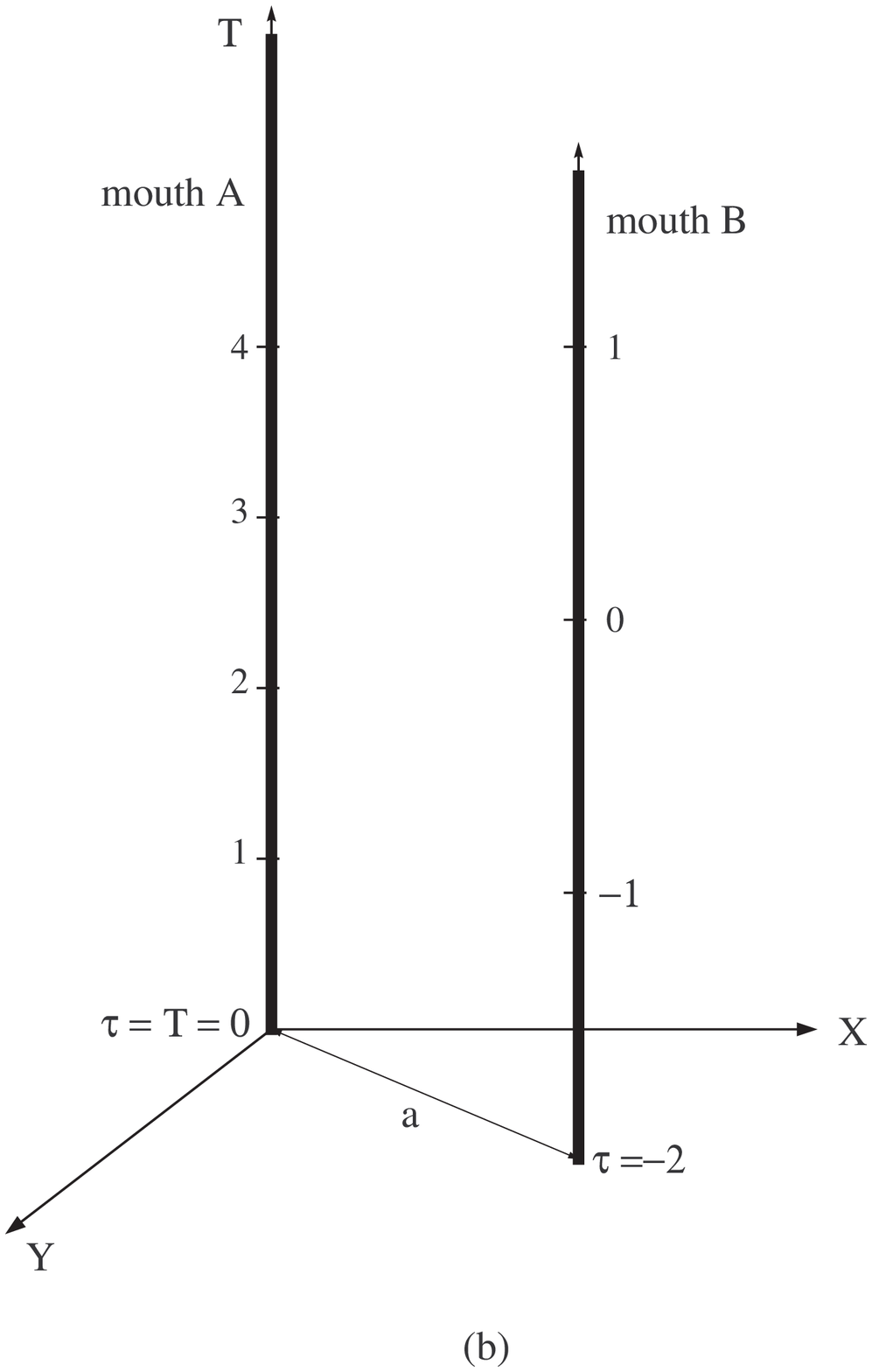}
  \caption[Wormhole spacetimes with closed timelike curves]
  {Depicted are two examples of wormhole spacetimes with closed
  timelike curves. The wormholes tunnels are arbitrarily short, and
  its two mouths move along two world tubes depicted as thick
  lines in the figure. Proper time $\tau$ at the wormhole throat is
  marked off, and note that identical values are the same event as seen
  through the wormhole handle. In Figure $(a)$, mouth $A$ remains at rest,
  while mouth $B$ accelerates from $A$ at a high velocity, then
  returns to its starting point at rest. A time shift is induced
  between both mouths, due to the time dilation effects of special
  relativity. The light cone-like hypersurface ${\it H}$ shown is
  a Cauchy horizon. Through every event to the future of ${\it H}$
  there exist CTCs, and on the other hand there are no CTCs to the past
  of ${\it H}$. In Figure $(b)$, a time shift between both mouths
  is induced by placing mouth $B$ in strong gravitational field.
  See text for details.}
  \label{fig:WH-time-machine}
\end{figure}

To create a time shift using the twin paradox analogue, consider
that the mouths of the wormhole may be moving one with respect to
the other in external space, without significant changes of the
internal geometry of the handle. For simplicity, consider that one
of the mouths $A$ is at rest in an inertial frame, whilst the
other mouth $B$, initially at rest practically close by to $A$,
starts to move out with a high velocity, then returns to its
starting point. Due to the Lorentz time contraction, the time
interval between these two events, $\Delta T_B$, measured by a
clock comoving with $B$ can be made to be significantly shorter
than the time interval between the same two events, $\Delta T_A$,
as measured by a clock resting at $A$. Thus, the clock that has
moved has been slowed by $\Delta T_A-\Delta T_B$ relative to the
standard inertial clock. Note that the tunnel (handle), between
$A$ and $B$ remains practically unchanged, so that an observer
comparing the time of the clocks through the handle will measure
an identical time, as the mouths are at rest with respect to one
another. However, by comparing the time of the clocks in external
space, he will verify that their time shift is precisely $\Delta
T_A-\Delta T_B$, as both mouths are in different reference frames,
frames that moved with high velocities with respect to one
another. Now, consider an observer starting off from $A$ at an
instant $T_0$, measured by the clock stationed at $A$. He makes
his way to $B$ in external space and enters the tunnel from $B$.
Consider, for simplicity, that the trip through the wormhole
tunnel is instantaneous. He then exits from the wormhole mouth $A$
into external space at the instant $T_0-(\Delta T_A-\Delta T_B)$
as measured by a clock positioned at $A$. His arrival at $A$
precedes his departure, and the wormhole has been converted into a
time machine. See Figure \ref{fig:WH-time-machine}.

For concreteness, following the Morris {\it et al}
analysis~\cite{mty}, consider the metric of the accelerating
wormhole given by
\begin{equation}
ds^2=-(1+glF(l)\cos\theta)^2\;e^{2\Phi(l)}\;dt^2+dl^2+r^2(l)\,(d\theta^2
+\sin^2\theta\,d\phi^2)   \,,
    \label{accerelatedWH}
\end{equation}
where the proper radial distance, $dl=(1-b/r)^{-1/2}\,dr$, is
used. $F(l)$ is a form function that vanishes at the wormhole
mouth $A$, at $l\leq 0$, rising smoothly from 0 to 1, as one moves
to mouth $B$; $g=g(t)$ is the acceleration of mouth $B$ as
measured in its own asymptotic rest frame. Consider that the
external metric to the respective wormhole mouths is $ds^2 \cong
-dT^2+dX^2+dY^2+dZ^2$. Thus, the transformation from the wormhole
mouth coordinates to the external Lorentz coordinates is given by
\begin{equation}
T=t\,, \qquad Z=Z_A+l\,\cos\theta\,, \qquad
X=l\,\sin\theta\,\cos\phi \,, \qquad  X=l\,\sin\theta\,\sin\phi
\,,
\end{equation}
for mouth $A$, where $Z_A$ is the time-independent $Z$ location of
the wormhole mouth $A$, and
\begin{equation}
T=T_B+v\gamma \,l\,\cos\theta\,, \qquad
Z=Z_B+\gamma\,l\,\cos\theta\,, \qquad X=l\,\sin\theta\,\cos\phi
\,, \qquad X=l\,\sin\theta\,\sin\phi \,,
\end{equation}
for the accelerating wormhole mouth $B$. The world line of the
center of mouth $B$ is given by $Z=Z_B(t)$ and $T=T_B(t)$ with
$ds^2=dT_B^2-dZ_B^2$; $v(t)\equiv dZ_B/dT_B$ is the velocity of
mouth $B$ and $\gamma=(1-v^2)^{-1/2}$ the respective Lorentz
factor; the acceleration appearing in the wormhole metric is given
$g(t)=\gamma^2\;dv/dt$~\cite{Misner}.

Novikov considered other variants of inducing a time shift through
the time dilation effects in special relativity, by using a
modified form of the metric (\ref{accerelatedWH}), and by
considering a circular motion of one of the mouths with respect to
the other~\cite{Novikov-CTCWH}. Another interesting manner to
induce a time shift between both mouths is simply to place one of
the mouths in a strong external gravitational field, so that times
slows down in the respective mouth. The time shift will be given
by
$T=\int_{i}^{f}\,(\sqrt{g_{tt}(x_A)}-\sqrt{g_{tt}(x_A)}\;)\;dt$
~\cite{Visser,frolovnovikovTM}.

\subsection{The Ori-Soen time machine}

A time-machine model was also proposed by Amos Ori and Yoav Soen
which significantly ameliorates the conditions of the EFE's
solutions which generate CTCs
\cite{Ori,Soen-Ori1,Soen-Ori2,Olum-SoenOriTM}. The Ori-Soen model
presents some notable features. It was verified that CTCs evolve, within a bounded region of space, from a well-defined initial slice $S$, a partial Cauchy surface, which does not display causality violation. The partial Cauchy surface and spacetime are asymptotically flat, contrary to the Gott
spacetime, and topologically trivial, contrary to the wormhole
solutions. The causality violation region is constrained within a
bounded region of space, and not in infinity as in the Gott
solution. The WEC is satisfied up until and beyond a time slice
$t=1/a$, on which the CTCs appear.

More recently, Ori presented a class of curved-spacetime vacuum solutions which develop closed timelike curves at some particular moment \cite{Ori:2005ht}. These vacuum solutions were then used to construct a time-machine model. The causality violation occurs inside an empty torus, which constitutes the time-machine core. The matter field surrounding this empty torus satisfies the weak, dominant, and strong energy conditions. The model is regular, asymptotically flat, and topologically trivial, although stability still remains the main open question.

\subsection{Warp drive and closed timelike curves}

Within the framework of general relativity, it is possible to warp
spacetime in a small {\it bubblelike} region \cite{Alcubierre}, in
such a way that the bubble may attain arbitrarily large
velocities, $v(t)$. Inspired in the inflationary phase of the
early Universe, the enormous speed of separation arises from the
expansion of spacetime itself. The model for hyperfast travel is
to create a local distortion of spacetime, producing an expansion
behind the bubble, and an opposite contraction ahead of it (see also \cite{warp}).

In the Alcubierre warp drive the spacetime metric is
\begin{equation}
d s^2=-d t^2+d x^2+d y^2+\left[d z-v(t)\;f(x,y,z-z_0(t))\; d
t\right]^2 \label{Cartesianwarpmetric}\,.
\end{equation}
The form function $f(x,y,z)$ possesses the general features of
having the value $f=0$ in the exterior and $f=1$ in the interior
of the bubble.  The general class of form functions, $f(x,y,z)$,
chosen by Alcubierre was spherically symmetric: $f(r)$ with
$r=\sqrt{x^2+y^2+z^2}$. Then
\begin{equation}
f(x,y,z-z_0(t)) = f(r(t)) \label{Alcubierreformfunction} \qquad
\hbox{with} \qquad
r(t)=\left\{[(z-z_{0}(t)]^2+x^2+y^2\right\}^{1/2}.
\end{equation}

Consider the following form
\begin{equation}
f(r)=\frac{\tanh\left[\sigma(r+R)\right]
-\tanh\left[\sigma(r-R)\right]}{2\tanh(\sigma R)}\,,
\label{E:form}
\end{equation}
in which $R>0$ and $\sigma>0$ are two arbitrary parameters. $R$ is
the ``radius'' of the warp-bubble, and $\sigma$ can be interpreted
as being inversely proportional to the bubble wall thickness. If
$\sigma$ is large, the form function rapidly approaches a {\it top
hat} function, i.e.,
\begin{equation}
\lim_{\sigma \rightarrow \infty} f(r)=\left\{ \begin{array}{ll}
1, & {\rm if}\; r\in[0,R],\\
0, & {\rm if}\; r\in(R,\infty).
\end{array}
\right.
\end{equation}

It can be shown that observers with the four velocity
\begin{equation}
U^{\mu}=\left(1,0,0,vf\right), \qquad\qquad
U_{\mu}=\left(-1,0,0,0\right).
\end{equation}
move along geodesics, as their $4$-acceleration is zero,
\emph{i.e.}, $a^{\mu} = U^{\nu}\; U^{\mu}{}_{;\nu}=0$. The spaceship, which in the original formulation is treated as a test particle which moves along the curve $z=z_0(t)$, can easily be seen to always move along a timelike curve, regardless of the value of $v(t)$. One can also verify that the proper time along this curve equals the coordinate time, by simply substituting $z=z_0(t)$ in Eq. (\ref{Cartesianwarpmetric}). This reduces to $d\tau=d t$, taking
into account $d x=d y=0$ and $f(0)=1$.

\begin{figure}
\centering
\includegraphics[width=3.4in]{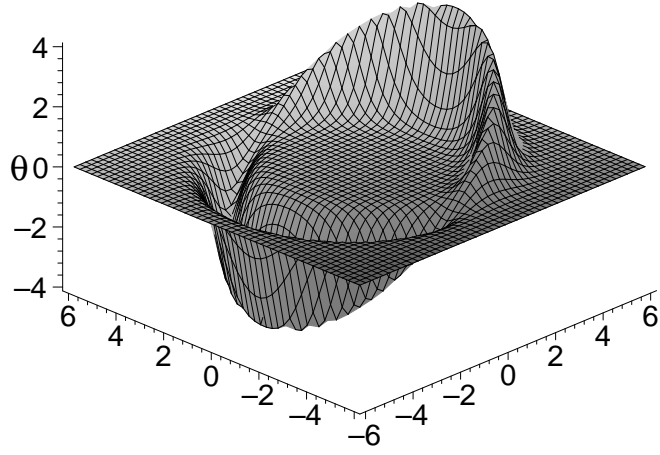}
\caption[The expansion of the volume elements for the Alcubierre
warp drive]{The expansion of the volume elements. These are
expanding behind the spaceship, and contracting in front of
it.}\label{Alcubierre-expansion}
\end{figure}

Consider a spaceship placed within the Alcubierre warp bubble. The
expansion of the volume elements, $\theta=U^{\mu}{}_{;\mu}$, is
given by $\theta=v\;\left({\partial f}/{\partial z} \right)$.
Taking into account Eq. (\ref{E:form}), we have (for Alcubierre's
version of the warp bubble)
\begin{equation}
\theta=v\;\frac{z-z_0}{r}\;\frac{d f(r)}{d r}.
\end{equation}
The center of the perturbation corresponds to the spaceship's
position $z_0(t)$. The volume elements are expanding behind the
spaceship, and contracting in front of it, as shown in Figure
\ref{Alcubierre-expansion}.

One may consider a hypothetical spaceship immersed within the
bubble, moving along a timelike curve, regardless of the value of
$v(t)$. Due to the arbitrary value of the warp bubble velocity,
the metric of the warp drive permits superluminal travel, which
raises the possibility of the existence of CTCs. Although the
solution deduced by Alcubierre by itself does not possess CTCs,
Everett demonstrated that these are created by a simple
modification of the Alcubierre metric \cite{EverettCTC}, by applying
a similar analysis as in tachyons.

The modified metric takes the form
\begin{equation}
ds^2=-dt^2+dx^2+dy^2+(dz-vfdt)^2  \,,
   \label{modwarpmetric}
\end{equation}
with
\begin{eqnarray}
v(t)=\frac{dz_0(t)}{dt} \qquad {\rm and} \qquad
r(t)=[(z-z_{0})^2+(y-y_0)^2+z^2]^{1/2} \,.
\end{eqnarray}
The spacetime is flat in the
exterior of a warp bubble with radius $R$, but now in the modified
version is centered in $(0,y_0,z_0(t))$. The bubble moves with a
velocity $v$, on a trajectory parallel with the $z$-axis. One may
for simplicity consider the form function given by Eq.
(\ref{E:form}). We shall also impose that $y_0\gg R$, so that the
form function is negligible, i.e., $f(y_0)\approx 0$.

Now, consider two stars, $S_1$ and $S_2$, at rest in the
coordinate system of the metric (\ref{modwarpmetric}), and located
on the $z$-axis at $t=0$ and $t=D$, respectively. The metric along
the $z$-axis is Minkowskian as $y_0\gg R$. Therefore, a light beam
emitted at $S_1$, at $t=0$, moving along the $z$-axis with
$dz/dt=1$, arrives at $S_2$ at $t=D$. Suppose that the spaceship
initially starts off from $S_1$, with $v=0$, moving off to a
distance $y_0$ along the $y-$axis and neglecting the time it needs
to cover $y=0$ to $y=y_0$. At $y_0$, it is then subject to a
uniform acceleration, $a$, along the the $z-$axis for $0<z<D/2$,
and $-a$ for $D/2<z<D$. The spaceship will arrive at the spacetime
event $S_2$ with coordinates $z=D$ and $t=2\sqrt{D/a}\equiv T$.
Once again, the time required to travel from $y=y_0$ to $y=0$ is
negligible.

The separation between the two events, departure and arrival is
$D^2-T^2=D^2(1-4/(aD)$ and will be spatial if the following
condition is verified
\begin{equation}
a>\frac{4}{D}\,.
\end{equation}
In this case, the spaceship will arrive at $S_2$ before the light
beam, if the latter's trajectory is a straight line, and both
departures are simultaneous from $S_1$. Inertial observers
situated in the exterior of the spaceship, at $S_1$ and $S_2$,
will consider the spaceship's movement as superluminal, since the
distance $D$ is covered in an interval $T<D$. However, the
spaceship's wordline is contained within it's light cone. The
worldline of the spaceship is given by $z=vt$, while it's future
light cone is given by $z=(v\pm 1)t$. The latter relation can
easily be inferred from the null condition, $ds^2=0$.

Since the quadri-vector with components $(T,0,0,D)$ is spatial,
the temporal order of the events, departure and arrival, is not
well-defined. Introducing new coordinates, $(t',x',y',z')$,
obtained by a Lorentz transformation, with a boost $\beta$ along
the $z$-axis. The arrival at $S_2$ in the $(t',x',y',z')$
coordinates correspond to
\begin{equation}
T'=\gamma (2\sqrt{D/a}-\beta D)\,, \qquad Z'=\gamma
(D-2\sqrt{D/a})  \,,
\end{equation}
with $\gamma =(1-\beta ^2)^{-1/2}$. The events, departure and
arrival, will be simultaneous if $a=4/(\beta^2D)$. The arrival
will occur before the departure if $T'<0$, i.e.,
\begin{equation}
a>4/(\beta^2D)  \,.
    \label{spatialcharacter}
\end{equation}

The fact that the spaceship arrives at $S_2$ with $t'<0$, does not
by itself generate CTCs. Consider the metric, Eq.
(\ref{modwarpmetric}), substituting $z$ and $t$ by $\Delta
z'=z'-Z'$ and $\Delta t'=t'-T'$, respectively; $v(t)$ by $-v(t)$;
$a$ by $-a$; and $y_0$ by $-y_0$. This new metric describes a
spacetime in which an Alcubierre bubble is created at $t'=T'$,
which moves along $y=-y_0$ and $x=0$, from $S_1$ to $S_2$ with a
velocity $v'(t')$, and subject to an acceleration $a'$. For
observers at rest relatively to the coordinates $(t',x',y',z')$,
situated in the exterior of the second bubble, it is identical to
the bubble defined by metric, Eq. (\ref{modwarpmetric}), as it is
seen by inertial observers at rest at $S_1$ and $S_2$. The only
differences reside in a change of the origin, direction of
movement and possibly of the value of acceleration. The stars,
$S_1$ and $S_2$, are st rest in the coordinate system of the
metric, Eq. (\ref{modwarpmetric}), and in movement along the
negative direction of the $z$-axis with velocity $\beta$,
relatively to the coordinates $(t',x',y',z')$. The two coordinate
systems are equivalent due to the Lorentz invariance, so if the
first is physically realizable, then so is the second. In the new
metric, by analogy with Eq. (\ref{modwarpmetric}), we have
$d\tau=dt'$, i.e., the proper time of the observer, on board of
the spaceship, traveling in the center of the second bubble, is
equal to the time coordinate, $t'$. The spaceship will arrive at
$S_1$ in the temporal and spatial intervals given by $\Delta t'>0$
and $\Delta z'<0$, respectively. As in the analysis of the first
bubble, the separation between the departure, at $S_2$, and the
arrival $S_1$, will be spatial if the analogous relationship of
Eq.(\ref{spatialcharacter}) is verified. Therefore, the temporal
order between arrival and departure is also not well-defined. As
will be verified below, when $z$ and $z'$ decrease and $t'$
increases, $t$ will decrease and a spaceship will arrive at $S_1$
at $t<T$. In fact, one may prove that it may arrive at $t<0$.

Since the objective is to verify the appearance of CTCs, in
principle, one may proceed with some approximations. For
simplicity, consider that $a$ and $a'$, and consequently $v$ and
$v'$ are enormous, so that $T\ll D$ and $\Delta t'\ll -\Delta z'$.
In this limit, we obtain the approximation $T\approx 0$, i.e., the
journey of the first bubble from $S_1$ to $S_2$ is approximately
instantaneous. Consequently, taking into account the Lorentz
transformation, we have $Z'\approx \gamma D$ and $T'\approx
-\gamma \beta D$. To determine $T_1$, which corresponds to the
second bubble at $S_1$, consider the following considerations:
since the acceleration is enormous, we have $\Delta t'\approx 0$
and $\Delta t=T_1-T\approx T_1$, therefore $\Delta z=-D\approx
\gamma \Delta z'$ and $\Delta t\approx \gamma \beta \Delta z'$,
from which one concludes that
\begin{equation}
T_1\approx -\beta D<0  \,.
\end{equation}

\subsection{The Krasnikov tube and closed timelike curves}

Krasnikov discovered an interesting feature of the warp drive, in
which an observer in the center of the bubble is causally
separated from the front edge of the bubble. Therefore he/she
cannot control the Alcubierre bubble on demand. Krasnikov proposed
a two-dimensional metric \cite{Krasnikov}, which was later
extended to a four-dimensional model \cite{Everett}. One Krasnikov tube in two dimensions does not generate CTCs. But the situation is quite
different in the 4-dimensional generalization, which we present for self-consistency and self-completeness.

Soon after the Krasnikov two-dimensional solution, Everett and
Roman~\cite{Everett} generalized the analysis to four dimensions,
denoting the solution as the {\it Krasnikov tube}. Consider that
the 4-dimensional modification of the metric begins along the path
of the spaceship, which is moving along the $x$-axis, occurring at
position $x$ at time $t \approx x$, the time of passage of the
spaceship. Also assume that the disturbance in the metric
propagates radially outward from the $x$-axis, so that causality
guarantees that at time $t$ the region in which the metric has
been modified cannot extend beyond $\rho = t - x$, where $\rho
={(y^2 + z^2)}^{1/2}$. The modification in the metric should also
not extend beyond some maximum radial distance $\rho_{max} \ll D$
from the $x$-axis. Thus, the metric in the 4-dimensional
spacetime, written in cylindrical coordinates, is given by
\cite{Everett}
\begin{equation}
ds^2=-dt^2+(1-k(t,x,\rho))dx dt+k(t,x,\rho)dx^2+d\rho^2+\rho^2
d\phi^2   \,,
   \label{4d-Krasnikov-metric}
\end{equation}
with
\begin{equation}
k(t,x,\rho)=1-(2-\delta)\theta_{\varepsilon}(\rho_{max}-\rho)
\theta_{\varepsilon}(t-x-\rho)[\theta_{\varepsilon}(x)-
\theta_{\varepsilon}(x+\varepsilon-D)]   \,.
    \label{4d:form}
\end{equation}
For $t\gg D+\rho_{max}$ one has a tube of radius $\rho_{max}$
centered on the $x$-axis, within which the metric has been
modified. This structure is denoted by the {\it Krasnikov tube}.
In contrast with the Alcubierre spacetime metric, the metric of
the Krasnikov tube is static once it has been created.

The stress-energy tensor element $T_{tt}$ given by
\begin{equation}
T_{tt}=\frac{1}{32
\pi(1+k)^2}\left[-\frac{4(1+k)}{\rho}\frac{\partial k}{\partial
\rho}+3\left(\frac{\partial k}{\partial
\rho}\right)^2-4(1+k)\frac{\partial^2 k}{\partial \rho^2}\right]
\,,
\end{equation}
can be shown to be the energy density measured by a static
observer \cite{Everett}, and violates the WEC in a certain range
of $\rho$, i.e., $T_{\mu\nu}U^{\mu}U^{\nu}<0$.

To verify the violation of the WEC, consider the energy density in
the middle of the tube and at a time long after it's formation,
i.e., $x=D/2$ and $t\gg x+\rho +\varepsilon $, respectively. In
this region we have $\theta_{\varepsilon}(x)=1$,
$\theta_{\varepsilon}(x+\varepsilon-D)=0$ and
$\theta_{\varepsilon}(t-x-\rho)=1$. With this simplification the
form function, Eq. (\ref{4d:form}), reduces to
\begin{equation}
k(t,x,\rho)=1-(2-\delta)\theta_{\varepsilon}(\rho_{max}-\rho)  \,.
     \label{4d:midtube-form}
\end{equation}
Consider the following specific form for
$\theta_{\varepsilon}(\xi)$ \cite{Everett} given by
\begin{equation}
\theta_{\varepsilon}(\xi)=\frac{1}{2}\left \{ \tanh \left
[2\left(\frac{2\xi}{\varepsilon}-1 \right )\right]+1 \right \} \,,
\end{equation}
so that the form function of Eq. (\ref{4d:midtube-form}) is
provided by
\begin{equation}
k=1-\left (1-\frac{\delta}{2}\right)\left \{ \tanh \left
[2\left(\frac{2\xi}{\varepsilon}-1 \right )\right]+1 \right \} \,.
\end{equation}

Choosing the following values for the parameters: $\delta =0.1$,
$\varepsilon =1$ and $\rho_{max}=100\varepsilon =100$, it can be shown that the negative character of the energy density is manifest in the
immediate inner vicinity of the tube wall.

Now, using two such tubes it is a simple matter, in principle, to generate CTCs. The analysis is similar to that of the warp drive, so that it will be
treated in summary.

Imagine a spaceship traveling along the $x$-axis, departing from
a star, $S_1$, at $t=0$, and arriving at a distant star, $S_2$, at
$t=D$. An observer on board of the spaceship constructs a
Krasnikov tube along the trajectory. It is possible for the
observer to return to $S_1$, traveling along a parallel line to
the $x$-axis, situated at a distance $\rho_0$, so that $D\gg
\rho_0\gg 2\rho_{max}$, in the exterior of the first tube. On the
return trip, the observer constructs a second tube, analogous to
the first, but in the opposite direction, i.e., the metric of the
second tube is obtained substituting $x$ and $t$, for $X=D-x$ and
$T=t-D$, respectively in Eq. (\ref{4d-Krasnikov-metric}). The
fundamental point to note is that in three spatial dimensions it
is possible to construct a system of two non-overlapping tube
separated by a distance $\rho_0$.

After the construction of the system, an observer may initiate a
journey, departing from $S_1$, at $x=0$ and $t=2D$. One is only
interested in the appearance of CTCs in principle, therefore the
following simplifications are imposed: $\delta$ and $\varepsilon$
are infinitesimal, and the time to travel between the tubes is
negligible. For simplicity, consider the velocity of propagation
close to that of light speed. Using the second tube, arriving at
$S_2$ at $x=D$ and $t=D$, then travelling through the first tube,
the observer arrives at $S_1$ at $t=0$. The spaceship has
completed a CTC, arriving at $S_1$ before it's departure.

\section{Discussion}

GTR has been an extremely successful theory, with a well
established experimental footing, at least for weak gravitational
fields. It's predictions range from the existence of black holes,
gravitational radiation to the cosmological models, predicting a
primordial beginning, namely the big-bang. However, it was seen
that it is possible to find solutions to the EFEs, with certain
ease, which generate CTCs. This implies that if we consider GTR
valid, we need to include the {\it possibility} of time travel in
the form of CTCs. A typical reaction is to exclude time travel due
to the associated paradoxes. But the paradoxes do not prove that
time travel is mathematically or physically impossible. Consistent
mathematical solutions to the EFEs have been found, based on
plausible physical processes. What they do seem to indicate is
that local information in spacetimes containing CTCs are
restricted in unfamiliar ways.

The grandfather paradox, without doubt, does indicate some strange
aspects of spacetimes that contain CTCs. It is logically
inconsistent that the time traveler murders his grandfather. But,
one can ask, what exactly impeded him from accomplishing his
murderous act if he had ample opportunities and the free-will to
do so. It seems that certain conditions in local events are to be
fulfilled, for the solution to be globally self-consistent. These
conditions are denominated {\it consistency constraints}
\cite{Earman}. To eliminate the problem of free-will, mechanical
systems were developed as not to convey the associated
philosophical speculations on free-will
\cite{Echeverria,NovikovCTC}. Much has been written on two
possible remedies to the paradoxes, namely the Principle of
Self-Consistency \cite{wh-timemachine,NovikovCTC,Carlini1,Carlini2} and
the Chronology Protection Conjecture
\cite{hawking}.

One current of thought, led by Igor Novikov, is the Principle of
Self-Consistency, which stipulates that events on a CTC are
self-consistent, i.e., events influence one another along the
curve in a cyclic and self-consistent way. In the presence of CTCs
the distinction between past and future events are ambiguous, and
the definitions considered in the causal structure of well-behaved
spacetimes break down. What is important to note is that events in
the future can influence, but cannot change, events in the past.
The Principle of Self-Consistency permits one to construct local
solutions of the laws of physics, only if these can be prolonged
to a unique global solution, defined throughout non-singular
regions of spacetime. Therefore, according to this principle, the
only solutions of the laws of physics that are allowed locally,
reinforced by the consistency constraints, are those which are
globally self-consistent.

Hawking's Chronology Protection Conjecture \cite{hawking} is a
more conservative way of dealing with the paradoxes. Hawking notes
the strong experimental evidence in favor of the conjecture from
the fact that "we have not been invaded by hordes of tourists from
the future". An analysis reveals that the value of the
renormalized expectation quantum stress-energy tensor diverges in
the imminence of the formation of CTCs. This conjecture permits
the existence of traversable wormholes, but prohibits the
appearance of CTCs. The transformation of a wormhole into a time
machine results in enormous effects of the vacuum polarization,
which destroys it's internal structure before attaining the Planck
scale. Nevertheless, Li has shown given an example of a spacetime
containing a time machine that might be stable against vacuum
fluctuations of matter fields~\cite{Li}, implying that Hawking's
suggestion that the vacuum fluctuations of quantum fields acting
as a chronology protection might break down. There is no
convincing demonstration of the Chronology Protection Conjecture,
but the hope exists that a future theory of quantum gravity may
prohibit CTCs.

Visser still considers the possibility of two other conjectures
\cite{Visser}. The first is the radical reformulation of physics
conjecture, in which one abandons the causal structure of the laws
of physics and allows, without restriction, time travel,
reformulating physics from the ground up. The second is the boring
physics conjecture, in which one simply ceases to consider the
solutions to the EFEs generating CTCs. Perhaps an eventual quantum
gravity theory will provide us with the answers. But, as stated by
Thorne \cite{thorneGRG13}, it is by extending the theory to it's
extreme predictions that one can get important insights to it's
limitations, and probably ways to overcome them. Therefore, time
travel in the form of CTCs, is more than a justification for
theoretical speculation, it is a conceptual tool and an
epistemological instrument to probe the deepest levels of GTR and
extract clarifying views.

%-----------------------------------
\section*{Acknowledgements}
FSNL acknowledges partial financial support of the
Funda\c{c}\~{a}o para a Ci\^{e}ncia e Tecnologia through the
grants PTDC/FIS/102742/2008 and CERN/FP/109381/2009.
%-----------------------------------

%-----------------------------------------------

%-----------------------------------------------

%-----------------------------------------------
\end{document}